\documentclass[useAMS,usenatbib,referee]{mn2e}
\usepackage{epsfig}
\usepackage{graphicx}

\begin{document}
\voffset=-0.5 in


\title[Magnetars origin and progenitors with enhanced rotation]
{Magnetars origin and progenitors with enhanced rotation}

\author[S.B. Popov, M.E. Prokhorov]{S.B. Popov$^{1}$ and M.E. Prokhorov$^{1}$ 
\thanks{E-mail: polar@sai.msu.ru (SBP); mike@sai.msu.ru (MEP)}\\
$^1${\sl Sternberg Astronomical Institute, Universitetski pr. 13, 
Moscow 119992, Russia}  } 
\date{Accepted ......  Received ......; in original form ......
      }

\maketitle

\begin{abstract}
Among a dozen known magnetar-candidates there are no binary objects.
As an estimate of a fraction of binary neutron stars is about 3-10\%
it is reasonable to address the question  of solitarity of magnetars,  to
estimate theoretically the fraction of binary objects among them, and to mark out the most
probable companions. 
We present  population synthesis calculations of binary 
systems. In this study we assume the hypothesis that magnetic field of a
magnetar is generated at the protoneutron star stage due to dynamo
mechanism, so rapid rotation of the core of a progenitor star is essential.
Our goal is to estimate 
the number of neutron stars originated from progenitors
with enhanced rotation.
In our calculations 
the fraction of neutron stars originated from such progenitors
is about 8-9\%. This should be considered as an upper limit to the fraction
of magnetars as some of progenitors can loose momentum.
Most of these objects 
are isolated due to coalescences of components prior to a
neutron star formation, or due to a system disruption after the second 
supernova explosion.
The fraction of such neutron stars 
in survived binaries is about 1\% or lower.
Their most numerous companions  are black holes.
  
\end{abstract}



\section{Introduction}

 There are more than 10 anomalous X-ray pulsars (AXPs) and 
soft gamma repeaters (SGRs) in our Galaxy known at the moment \citep{wt2004}.
The standard model of AXPs and SGRs involves highly magnetized neutron stars
-- magnetars \citep{td1993}.
All these sources are isolated, ie. no binary components were reported (we
can add also radio pulsars with large magnetic fields, all of them are
single objects too). 
Population synthesis studies suggest that  from few up to $\sim 10$\% of 
neutron stars (NSs) are expected
to have  companions (see, for example, \citet{it1996} and calculations below). 
This number depends on the kick velocity distribution and on the fraction of
binaries in the whole stellar population.
Probably, now it is reasonable to start a discussion 
{\it "Why all known magnetars are isolated?"}.\footnote{ We do not discuss 
here alternatives to the magnetar scenario, see, for example, \citet{mph2001}.}

Immediately two possibilities arise:

\begin{itemize}
\item {\it Large kick velocities.}
It was suggested that magnetars have large kicks.
This option naturally explains the solitarity of magnetars: 
if a progenitor star was a member of a binary then anyway in most cases the
system should be disrupted. 
So, if all magnetars are born with very large recoil
velocities then they have to be isolated.
\item {\it Particular evolutionary paths.}
Magnetars seldom can have companions even for standard kicks. 
This can happen if magnetars are mainly born from single stars, or if due to
some properties of evolution 
nearly all progenitor binaries are disrupted prior or immediately
after a magnetar formation.
\end{itemize}

The first option is quite possible and widely discussed (see, for example,
\citet{whw2000} and references therein). 
However, let us focus here on the second possibility connected with
evolution in a binary as actually there is no direct evidence for large
kicks of magnetars.
In particular, we start with a hypothesis that magnetars are born only from
stars with enhanced rotation, and then we discuss evolutionary paths that
produce such stars.
 
\citet{td1993} and \citet{tm2001} demonstrate that
for the generation of a magnetar-scale magnetic field 
rapid rotation of a newborn NS 
is absolutely necessary. Rotation  of massive stars is now actively
studied in connection with gamma-ray bursts (GRBs) \citep{hetal2003, 
letal2003, tcq2004, fh2005}.    
\citet{hetal2003} 
suggest that isolated stars do not produce cores with the rate of rotation
sufficient to launch a GRB. 
In this sense it is reasonable to expect that GRBs are formed in binaries. 
This topic was discussed, for example, by 
\citet{letal2003}. These authors show that primary (ie. initially more
massive) components loose angular momentum via a stellar wind;
only secondary components of massive binaries produce cores with rapid
rotation. 

In the case of a NS formation \citet{hetal2003} show that rapid rotation can
be reached if no additional loss of angular momentum (suggested by
\citet{s2002}) occur. However, inclusion of such losses 
significantly decreases  spins
of stellar cores. Probably, in this case an additional enhancement of rotation
is necessary. As in the case of GRB progenitors extra-momentum can be gained
in course of binary evolution. Without detailed calculations it is possible
to point at two possibilities: coalescence of binary components and spin-up
of a secondary component via Roche lobe overflow.
In the latter case systems like Be/X-ray binaries 
can be considered as an example of progenitors of magnetars. 
Most probably such  systems are unbounded after 
the second supernova (SN) explosion. 
It can be a natural explanation for solitary nature of magnetars. 
Possibly in rare cases a magnetar can have as a component
a NS or a black hole (BH).    
If coalescences are the main producers of progenitors of magnetars 
then the resulting compact objects are single by definition. 
Still, a primary component also can gain angular momentum due to
synchronization of its spin with orbital rotation
if the mass transfer is accompanied by a common envelope
formation and if the separation between two components significantly
decreases. 

Obviously,  a 
detailed population synthesis of close binaries is necessary to determine
the fraction of stars with significant angular momentum at a pre-SN stage.
In the following section we present such calculations.


\section{Calculations and results}

 In this short note we use the ``Scenario Machine'' code for binary
population synthesis \citep{lpp1996} to calculate the number of NSs born from
progenitors with enhanced 
rotation.\footnote{Online material is available at
 http://xray.sai.msu.ru/sciwork/scenario.html and
http://xray.sai.msu.ru/~$\sim$mystery/articles/review/.
The first link is to the web-demonstrator of the program, only single track
calculations are available.
The second URL is for the online version of \citep{lpp1996}.}
The code is developed at the Sternberg Astronomical Institute since early
80s. Initially it was 
mainly designed to calculate evolution of massive binaries. 
The main feature is
the detailed calculation of magneto-rotational evolution of NSs. 
During more than 20 years the code was applied to many problems of binary
evolution. Topics studied before 1996 (including detailed studies of
coalescence rate of binary NSs and black holes) are summurized in
\citep{lpp1996}. In the last 10 years the code was also used to study
evolution of SN rates, populations of binaries after bursts of starformation,
population of binary radio pulsars with massive companions, Be/X-ray
systems, and evolution of binaries in globular clusters. 

 Among all possible evolutionary paths  that end in
formation of a NS we select those that lead to angular momentum increase of
a progenitor. These paths include:

\begin{itemize}
\item Coalescence prior to a NS formation.
\item Roche lobe overflow by a primary (it spins-up the secondary
companion). 
\item Roche lobe overflow by a primary with a common envelope (it spins-up
the primary if there is enough time for synchronization).
\item Roche lobe overflow by a secondary without a common envelope
formation prior to an explosion of a primary (it spins-up a primary).          
\item Roche lobe overflow by a secondary  with a common envelope (it spins-up    
the secondary if there is enough time for synchronization).
\end{itemize}
Below we dub NSs originated through these paths as ``magnetars'' (of course,
if they really have strong magnetic fields should be a subject of a separate
detailed study, so here it is just a hypothesis; also we neglect a
possibility of significant angular momentum losses after rotation has been
enhanced by some processes).

The question of synchronization needs few additional comments.
We consider that the synchronization is possible if for a survived binary
the duration of the stage which follows the common envelope one is
twice longer than $T_{\mathrm{sync}}$ (see, for example, \citet{s1994} p. 49): 

$$
\begin{array}{c}
 T_{\mathrm{sync}}=3500\, \left[(1+q)/q\right]
\left(L/L_{\sun}\right)^{-1/4}  \times \\
 \left(M/M_{\sun}\right)^{1.25} 
\left(R/R_{\sun}\right)^{-3} 
\left(P_\mathrm{orb}/{\mathrm{days}}\right)^{2.75}\,\,\, {\rm yrs}.
\\
\end{array}
$$

 Potentially, if  a primary had
transfered a significant angular momentum
onto a secondary, but later on the secondary, in its turn, filled its Roche
lobe and started to transfer momentum back, then the secondary's rotation is
reduced. Such tracks appear in our calculations.
We save for analysis tracks for which a secondary component forms a common
envelope and synchronizes with orbital rotation. We exclude tracks for
which a secondary only filled its Roche lobe (ie. no synchronization) 
and transfered angular momentum to the primary.

We studied three variants of the kick velocity distribution.
 In the first one
kick velocities are assumed to have the bimodal velocity distribution
\citep{acc2002}.  We consider it as the main variant and present the results
for it in more details below. 
For comparison we make calculations for two other distributions.
At first  we run a variant with significantly lower average velocity.
We use single-mode maxwellian with the
most probable velocity 127 km~s$^{-1}$ (it corresponds to the first peak in 
the \citet{acc2002} distribution). Then we produce calculations for 
higher kick velocities. We accept the
distribution proposed by \citet{h2005}. It is a single maxwellian
distribution with the most probable velocity $\sim 370$ km~s$^{-1}$.  In
this distribution there are less NSs with small ($\la 200$  km~s$^{-1}$) and
large ($\ga 700$  km~s$^{-1}$) in comparison with the bimodal distribution.  

We neglect any dependences of a kick velocity on any
parameters, including an evolution in a binary. The latter effect can be
significant \citep{petal2004}, however, at the moment no good theory or
phenomenology of it exists, so we neglect it. Black hole kicks are assumed
to be zero. The critical mass which separates BH progenitors from NS
progenitors is assumed to be $35\,M_{\sun}$ at the main sequence.

We run the code for two values of the parameter 
$ \alpha_q$ which characterize the mass ratio distribution of components,
$f(q)$, where $q$ is the mass ratio. At first, the mass of a primary is taken
from a Salpeter distribution, and then the $q$ distribution is applied.

$$
 f(q)\propto q^{\alpha_q}, \,\,\, q=M_2/M_1, \, q\le 1
$$
 We use $\alpha_q=0$ 
(flat distribution, ie. all variants of mass ratio are equally
probable) and $\alpha_q=2$ (close masses are more probable, so  numbers of 
NS and BH progenitors are increased in comparison with $\alpha_q=0$).   

 The distribution of major semiaxes is assumed to be
flat in the logarithmic scale in the interval (10-$10^7$)~$R_{\sun}$.
We used two variants of mass loss rate by normal stars.
Moderate mass loss by normal stars was taken according to \citet{vetal1998},
high mass rate was calculated according to papers by the Geneva group (see
\citet{mm2000} and references therein).   As the reference variant we
consider the one with moderate mass loss, and detailed results (see Table~3)
are presented only for this case.
Other parameters are typical for the ``Scenario Machine''. 
Detailed description of the  `Scenario Machine'' and comparison with other
codes can be found in \cite{lpp1996}.

\begin{table}
 \centering
 \begin{minipage}{140mm}
  \caption{Results of calculations for moderate mass loss}
  \begin{tabular}{@{}lrrrrrr@{}}
  \hline
   Name     & \multicolumn{2}{c}{Bi--Maxwell}  & \multicolumn{2}{c}{Maxwell,
$V_p=127$~km/s}  & \multicolumn{2}{c}{Maxwell, $V_p=370$~km/s} \\
            & $\alpha_q=0$   &  $\alpha_q=2$ & $\alpha_q=0$   &
$\alpha_q=2$ & $\alpha_q=0$   &  $\alpha_q=2$ \\
 \hline
 Total number of tracks       & 100 000  &       100 000    & 100 000  & 100
000 & 100 000  & 100 000 \\
 Total number of NSs          & 113 805  &       126 698    & 109 857  & 128
205 & 113 442  & 133 085 \\
 Number of binary NSs         &   6 604  &         7 065    &  16 466  &  17
814 &   3 116  &   3 242 \\
 Fraction of binary NSs       &   3.1\%  &         3.1\%    &   7.8\%  &
7.8\% &   1.5\%  &   1.4\% \\
 Number of `magnetars''      &  18 369  &        20 494    &  16 884  &  18
096 &  18 629  &  20 875 \\
 Number of `binary magnetars'' &   114  &           208    &     397  &
307 &      84  &     145 \\
 Fraction of `magnetars''    &  8.6\%   &        9.0\%     &   8.0\%  &
7.9\% &   8.7\%  &   9.0\% \\
 From coalescence             & 60.1\%   &       35.7\%     &  65.4\%  &
40.4\% &  59.3\%  &  35.0\% \\
 From primary components      &  2.5\%   &        1.6\%     &   2.7\%  &
1.7\% &   2.4\%  &   1.5\% \\
 From secondary components    & 37.4\%   &       62.7\%     &  31.9\%  &
57.9\% &  38.3\%  &  63.5\% \\
\hline
\end{tabular}
\end{minipage}
\end{table}   

\begin{table}
 \centering
 \begin{minipage}{140mm}
  \caption{Results of calculations for strong mass loss}
  \begin{tabular}{@{}lrrrrrr@{}}
  \hline
   Name     & \multicolumn{2}{c}{Bi--Maxwell}  & \multicolumn{2}{c}{Maxwell,
$V_p=127$~km/s}  & \multicolumn{2}{c}{Maxwell, $V_p=370$~km/s} \\
            & $\alpha_q=0$   &  $\alpha_q=2$ & $\alpha_q=0$   &
$\alpha_q=2$ & $\alpha_q=0$   &  $\alpha_q=2$ \\
 \hline
 Total number of tracks       & 100 000  &       100 000    & 100 000  & 100
000 & 100 000  & 100 000 \\
 Total number of NSs          & 126 845  &       145 289    & 121 571  & 137
610 & 126 607  & 145 869 \\
 Number of binary NSs         &   8 303  &         9 101    &  20 217  &  22
516 &   4 020  &   4 359 \\
 Fraction of binary NSs       &   3.7\%  &         3.7\%    &   9.1\%  &
9.5\% &   1.8\%  &   1.8\% \\
 Number of `magnetars''      &  30 180  &        29 226    &  26 348  &  23
652 &  31 068  &  30 621 \\
 Number of `binary magnetars'' &   157  &           296    &     514  &
795 &     133  &     223 \\
 Fraction of `magnetars''    &  13.3\%  &        11.9\%    &  11.9\%  &
10.0\% &  13.7\%  &  12.5\% \\ 
 From coalescence             &  56.7\%  &        29.4\%    &  65.0\%  &
36.3\% &  55.1\%  &  28.1\% \\
 From primary components      &   0.7\%  &         1.8\%    &   0.8\%  &
2.2\% &   0.7\%  &   1.7\% \\
 From secondary components    &  42.6\%  &        68.8\%    &  34.2\%  &
61.5\% &  44.2\%  &  70.2\% \\
\hline
\end{tabular}
\end{minipage}
\end{table}

In all runs we calculate 100 000 tracks of massive binaries with minimum
mass of a primary 10~$M_{\sun}$ (at least one
of components produce a NS or a BH). Results are presented in 
Tables~1 and 2. The first table refers to moderate mass loss scenario, the
second -- to enhanced mass loss.
Six columns represent numbers for three kick velocity distribution
and two choices of $\alpha_q=2$.
To produce percentage of ``magnetars'' and fraction of binary NSs
we add 100 000 NSs which
should originate from single stars. Absolute numbers are related just to our
runs without any normalization. 

Fractions of binary NSs are significantly different for three variants of the
kick velocity distribution. For the lowest one we obtain that $\sim 15-17$  \%
of newborn NSs originated from binary systems remain
in bounded binaries (of course, most of the are from the
primary components). If in addition to binaries we include the same number
of isolated progenitors the fraction goes down to $\sim 8-9$ \%.
For \citet{acc2002} the fraction of binary newborn NSs is $\sim 6$ \%
(without inclusion of single progenitors). Finally, for the distribution
proposed by \citet{h2005} this fraction is $\sim 3$\%.  

Not surprisingly, the fraction of ``binary magnetars'', ie. NSs originated
from progenitors with enhanced rotation which remained bound with their
components after a SN explosion, is very low (about a fraction
of a percent from the number of ``magnetars''). 
This is due to the fact that
most of ``magnetars'' come from coalescence or from secondary components. In
the latter case the more massive star explodes, so more than 1/2 of the
system's mass is lost and there is a very high probability that the binary
disrupts. Also it is not a surprise, that for this small fraction of ``binary
magnetars'' the most probable components are BHs; they constitute
more than one half of companions. Of course, in all the cases when a BH is
one of companions, 
a ``magnetar'' is formed from the secondary component of a binary.
In general, if a NS is formed from a secondary spined-up
component then its companion is a compact object: a BH, a white dwarf, 
or another NS.
If a ``magnetar'' is formed from a primary component then the secondary is
usually at the stage of Roche overflow without common envelope; but the
number of such systems in nearly zero. 
We have to note, that nearly in all cases when a secondary produces a
``magnetar'' and a primary is a BH there is an episode of a common
envelope formation (with synchronization) by the secondary.

 Results are sensitive (but not very significantly) to assumptions about
mass loss by normal stars (compare Tables 1 and 2). For higher mass loss
(Table 2) fraction of magnetars is higher by $\sim$ 30~-~50 \%.

Fraction of ``magnetars'' does not depend significantly on kick. 
The main reason
is that more than 1/3 of them (up to 2/3) originate after coalescence of
normal stars, and the number of coalescence does not depend on kick at all.
The rest of ``magnetars'' originate from secondary companions. A system
usually survives after the first SN explosion even for significant (few
hundred km~s$^{-1}$) kicks, so again number of ``magnetars'' is not very
sensitive to kicks.

In the Table~3 we present detailed result for the bimodal kick distribution
and moderate mass loss.
We separate different paths and give percentage of ``magnetars'' and binary
``magnetars'' for each channel. Most important paths for NS formation from
progenitors with enhanced rotation are coalescence and single Roche lobe
overflow by a primary component without a common envelope formation. 

 
\begin{table}
 \centering
  \caption{Detalied results of calculations for the bimodal kick and
moderate mass loss}
  \begin{tabular}{@{}lrrrr@{}}
  \hline
   Track    & \multicolumn{2}{c}{$\alpha_q=0$}   &
\multicolumn{2}{c}{$\alpha_q=2$} \\
            & \multicolumn{1}{c}{all} & \multicolumn{1}{c}{binary magnetars}   
            & \multicolumn{1}{c}{all} & \multicolumn{1}{c}{binary magnetars}
\\
 \hline
 Merge                         &  60.1\% &   --- & 35.7\% &     --- \\
 Primary RLO with CE+Sync.     &   1.6\% & 0.2\% & 0.02\% &  0.01\% \\
 Secondary RLO with CE+Sync.   &   1.4\% &  0.3\%&  2.5\% &   0.8\% \\
 Two RLO w/o CE on Primary     &   9.4\% &   0\% & 19.9\% &     0\% \\
 Two RLO w/o CE on Secondary   &0.005\%  &   0\% & 0.04\% &     0\% \\
 Single RLO w/o CE on Primary  &  26.1\% & 0.07\%& 39.3\% &  0.08\% \\
 Single RLO w/o CE on Secondary&   0.7\% & 0.08\%&  1.4\% &  0.06\% \\
\hline
\end{tabular} 
\end{table}


\section{Discussion}

Here we briefly comment on several subjects connected with our study.

\subsection{Birth rate of magnetars}

\citet{ketal1998} suggest that magnetars are born with the rate about
one per 1000 years, i.e. about few percents of NSs are born with high
magnetic fields. Our estimates are in reasonable correspondence with this
hypothesis. If highly magnetized NSs are born only from rapidly
rotating progenitors then our numbers ($\sim 8-9$\%) 
are good upper limits as without any doubts not all massive stars 
with angular momenta enhanced during their evolution would produce magnetars. 


Note, that an actual value of angular momentum transfer was not
calculated by us, also momentum losses after a possible increase were not
carefully treated. 
We do not calculate if the momentum is transfered to the core of the
star or not.
Tracks just were  selected  using {\it qualitative} criteria.
Only for the case of coalescence we can be sure that angular momentum of the
core is significantly increased. 
However, even if only coalescences contribute to
formation of magnetars, the number of such NSs is high enough to be
consistent with the estimate by \citet{ketal1998}.

\subsection{Massive stars and magnetars}

Several authors suggested that magnetars should be born from the most 
massive stars that still could produce NSs
\citep{fetal2005, getal2005}. As it is well known, massive stars ($M\ga
20$-$25\, M_{\sun}$) extensively loose angular momentum via intensive
stellar winds. For such progenitors rotation enhancement in binaries seems
inevitable. Unless, as it is noted for example by \citet{tc2004}, stars with
$M\ga 50\, M_{\sun} $ do not loose angular momentum as they avoid the red
supergiant stage.  \citet{cetal2005} suggest  few arguments  to
show that the connection between magnetars and massive progenitors can be not
valid (at least in the case of SGR 1806-20). However, this paper was
strongly critisized by \cite{mg2005}, and the link between massive
progenitors and magnetar candidates seems to be solid.    
Recently \citet{muno2005} reported
the discovery of an anomalous X-ray pulsar with very massive progenitor
(($M\ga 40\, M_{\sun}$). A forty solar masses isolated star would loose its
angular momentum due to the stellar wind. This can be concidered as an
indirect argument in favor of evolution in a binary if one agrees with the
hypothesis about the magnetic field generation used in this paper. 

\subsection{Coalescence of two helium stars}

\citet{fh2005} suggested a scenario in  which a GRB progenitor is formed
after coalescence of two helium stars. They also discuss a possibility of a NS
formation via this path. We found such tracks in our calculations:  $\la 1$\%
of all ``magnetars'' (ie. about 0.1\% of all NSs) are formed in this way.
So the formation rate of such objects is about 1 in $\ga 10^5$ yrs. 
These values do not show strong dependence on mass loss rate and $\alpha_q$. 
We made calculations for both variants of mass loss (moderate and high) and
for $\alpha_q$ equal to 0 and 2: numbers are close to each other.
In principle, such sources with high magnetic fields and extreme rotation at
birth can produce spectacular manifestations. Similar objects can be formed
via coalescence of helium stars with white dwarfs or via coalescence of two
white dwarfs. However, we do not study this
possibility in details, and the obtained number should be considered as a
rough estimate as it depends on many parameters which we do not discuss here
(the critical mass which separates NSs from BHs, etc.).

As a by-product we estimate the rate of BH formation after
coalescence of two helium stars (no coalescence with white dwarf
participation produce BHs). 
%
In our calculations with parameters presented in this paper (bimodal kick,
two variants of mass loss and $\alpha_q$) we do not find any.
Changing evolutionary scenario we were able to obtain few.
It gives us the opportunity to put an upper limit for formation rate of BHs
after coalescence of two helium stars. It is $\la 5 \, 10^{-6}$~yr$^{-1}$.
Probably we slightly underestimate this number as some of coalescence which
is our calculations lead to a NS formation can result in a BH. However, this
can not be  a significant effect.
This value is too low to explain the rate of GRBs. 



\subsection{Velocity distribution}


Here we avoid discussing the velocity distribution of magnetars.
Basing on associations with SN remnants it was suggested that
magnetars can have large spatial velocities (see \citet{g2003} and
references therein). 
High velocities can appear not only due to large kicks, 
they can be potentially connected also with a disruption of a
binary system after an explosion of a secondary
companion (see the discussion of this subject in \citet{it1996}).  
Still, exact velocity determinations for SGRs and AXPs are very welcomed.

We also note that kick velocity was assumed to be uncorrelated with other
parameters. However, this is just an assumption. The generation of very large
magnetic fields by itself can produce an additional acceleration (due to
anisotropic neutrino losses, for example). Fast rotation of a core can lead
to fragmentation \citep{betal1988} and to high velocity of a compact object.
In addition a kick can potentially
influence rotation of a newborn NS \citep{pp1998,sp1998}.
So, correlations can be extremely important.
This subject should be studied separately in more details.

\section{Conclusions}

Our conclusions are the following. We made population synthesis of binary
systems to derive the relative number of NSs originated from progenitors
with enhanced rotation -- ``magnetars''. 
With an inclusion of single stars (with the total
number equal to the total number of binaries) the fraction of ``magnetars''
is $\sim$~8-9\%.
Most of these NSs are isolated due to coalescences of components prior to 
NS formation, or due to a system disruption after the second SN explosion.
The fraction of ``magnetars'' in survived binaries is about 1\% or lower.
The most numerous companions of ``magnetars'' are BHs.

\section*{Acknowledgments}

 We appreciate  discussions with Dr. A.V. Tutukov.
We want to the the referee for useful suggestions.
The work was supported by the RFBR grants 04-02-16720 and 03-02-16068. 
S.P. thanks the ``Dynasty'' Foundation (Russia).



\begin{thebibliography}{99}
\bibitem[\protect\citeauthoryear{Arzoumanian, Chernoff, Cordes}{2002}]{acc2002}
Arzoumanian Z., Chernoff D.F.,  Cordes J.M., 2002, ApJ, 568, 289
\bibitem[\protect\citeauthoryear{Berezinskii et al.}{1988}]{betal1988}
Berezinskii V. S., Castagnoli C., Dokuchaev V. I., Galeotti, P., 1988, 
Nuovo Cimento, 11C, 287
\bibitem[\protect\citeauthoryear{Cameron et al.}{2005}]{cetal2005}
Cameron P.B., Chandra P., Ray A., et al., 2005, Nature, 434, 1112
[astro-ph/0502428]
\bibitem[\protect\citeauthoryear{Figer et al.}{2005}]{fetal2005}
Figer D.F., Najarro F., Geballe T.R., Blum R.D., Kudritzki R.P., 2005, ApJ,
622, L49 [astro-ph/0501560] 
\bibitem[\protect\citeauthoryear{Fryer, Heger}{2005}]{fh2005}
Fryer C.L., Heger A., 2005, ApJ, 623, 302 [astro-ph/0412024]
\bibitem[\protect\citeauthoryear{Gaensler}{2004}]{g2003}
Gaensler B.M., 2004, Advances in Space Research, 33, 645 [astro-ph/0212086]
\bibitem[\protect\citeauthoryear{Gaensler et al.}{2005}]{getal2005}
Gaensler B.M.,  McClure-Griffiths N.M., Oey M.S. et al., 2005, ApJ,
620, L95 [astro-ph/0501563] 
\bibitem[\protect\citeauthoryear{Heger et al.}{2003}]{hetal2003}
Heger A., Woosley S.E., Langer N., Spruit H.C., 2003, in: Proc. of IAU Symp.
215 `Stellar evolution'',
ed. A. Maeder, P. Eenes, p. 591, ASP, San Francisco
[astro-ph/0301374]
\bibitem[\protect\citeauthoryear{Hobbs et al.}{2005}]{h2005}
Hobbs G.,  Lorimer D.R.,  Lyne A.G., Kramer M., 2005,
MNRAS 360, 963
\bibitem[\protect\citeauthoryear{Iben, Tutukov}{1996}]{it1996}
Iben  I., Tutukov A.V., 1996, ApJ, 456, 738
\bibitem[\protect\citeauthoryear{Kouveliotou et al.}{1998}]{ketal1998}
Kouveliotou C.,  Dieters S., Strohmayer T., et al., 1998, Nature, 393, 235
\bibitem[\protect\citeauthoryear{Langer et al.}{2003}]{letal2003}
Langer N., Yoon S.-C., Petrovic J., Heger A., 2003, 
astro-ph/0302232
\bibitem[\protect\citeauthoryear{Lipunov, Postnov, Prokhorov}{1996}]{lpp1996}
Lipunov V.M., Postnov K.A.,  Prokhorov M.E., 1996,
Astrophys. and Space Science Rev.,  9, 1
\bibitem[\protect\citeauthoryear{Maeder, Meynet}{2000}]{mm2000}
Maeder A., Meynet G., 2000, ARA\&A, 38, 143
\bibitem[\protect\citeauthoryear{McClure-Griffiths \&
Gaensler}{2005}]{mg2005}
McClure-Griffiths N.M., Gaensler B.M., 2005, ApJ, 630, L161 [astro-ph/0503171]   
\bibitem[\protect\citeauthoryear{Menou, Perna, Hernquist}{2001}]{mph2001}
Menou K., Perna R., Hernquist L., 2001, ApJ, 559, 1032
\bibitem[\protect\citeauthoryear{Muno et al.}{2005}]{muno2005}
Muno M.P.,  Clark, J.S. Crowther P.A. et al., 2005, astro-ph/0509408
\bibitem[\protect\citeauthoryear{Podsiadlowski et al.}{2004}]{petal2004}
Podsiadlowski Ph., Langer N., Poelarends A.J.T., Rappaport S.,
Heger A., Pfahl E., 2004, ApJ, 612, 1044
\bibitem[\protect\citeauthoryear{Postnov, Prokhorov}{1998}]{pp1998}
Postnov K. A., Prokhorov M. E., 1998, Astronomy Letters, 24, 568
\bibitem[\protect\citeauthoryear{Shore}{1994}]{s1994}
Shore S., 1994, in: ``Interacting binaries'', Saas-Fee Advanced Course 22,
ed. Nussbaumer H., Orr A.,  Berlin: Springer
\bibitem[\protect\citeauthoryear{Spruit}{2002}]{s2002}
Spruit H., 2002, A\&A, 381, 923 
\bibitem[\protect\citeauthoryear{Spruit, Phinney}{1998}]{sp1998}
Spruit H.C., Phinney E.S., 1998, Nature, 393, 139
\bibitem[\protect\citeauthoryear{Thompson, Duncan}{1993}]{td1993}
Thompson C., Duncan R.C., 1993, ApJ, 408, 194
\bibitem[\protect\citeauthoryear{Thompson, Murray}{2001}]{tm2001}
Thompson C., Murray N., 2001, ApJ, 560, 339
\bibitem[\protect\citeauthoryear{Thompson, Chang, Quataert}{2004}]{tcq2004}
Thompson T.A., Chang P., Quataert E., 2004, ApJ, 611, 380
\bibitem[\protect\citeauthoryear{Tutukov, Cherepashchuk}{2004}]{tc2004}
Tutukov A.V., Cherepashchuk A.M., 2004, Astronomy reports 48, 39
\bibitem[\protect\citeauthoryear{Vanbeveren et al.}{1998}]{vetal1998}
Vanbeveren D., de Donder E., van Bever J., van Rensbergen W., de Loore C.,
1998, New Astronomy,  3,  443
\bibitem[\protect\citeauthoryear{Wheeler et al.}{2000}]{whw2000}
Wheeler J.C., Yi I., Hoflich P., Wang L., 2000, ApJ,  537, 810
\bibitem[\protect\citeauthoryear{Woods, Thompson}{2004}]{wt2004} 
Woods P.M., Thompson C., 2004, astro-ph/0406133
\end{thebibliography}
\end{document}